\documentclass[aps,twocolumn,showpacs,floatfix]{revtex4-2}

\usepackage{amsfonts}
\usepackage{amsmath}
\usepackage{amssymb}
\usepackage{graphicx}
\usepackage{color}
\usepackage{verbatim}
\usepackage{lipsum}
\usepackage{dsfont}
\usepackage{ulem}

\usepackage[T1]{fontenc}

\usepackage{hyperref}
\usepackage{makecell}

\usepackage{physics}

\begin{document}

\title{Using NV centers in diamond to detect DC to very-low frequency magnetic fields}

\author{Valts Krumins}
\email{valts.krumins@lu.lv}
\author{Ivars Krastins}
\author{Oskars Rudzitis}
\author{Reinis Lazda}
\author{Florian Gahbauer}
\author{Marcis Auzinsh}

\affiliation{Laser Centre, University of Latvia, Jelgavas street 3, LV-1004, Riga, Latvia}

\pacs{76.30.Mi,76.70.Hb,75.10.Dg}
%76.30.Mi Color centers and other defects
%76.70.Hb Optically detected magnetic resonance (ODMR)
%75.10.Dg Crystal-field theory and spin Hamiltonians (see also 71.70.Ch Crystal and ligand fields)

\begin{abstract}

In this work we present a compact and portable tabletop magnetometer that utilizes negatively charged nitrogen-vacancy (NV) centers in diamond. The magnetometer is operated using a dual microwave resonance detection approach in combination with an optically detected magnetic resonance (ODMR) technique (mitigating drifts in results due to changes of the diamond temperature), capable of simultaneously exciting and registering two ODMR transitions. The experimentally measured magnetic field noise-floor is $\approx 2.3~\textrm{nT}\sqrt{\textrm{Hz}}$ while the calculated shot-noise-limited magnetic field sensitivity is $\approx 585~\textrm{pT}\sqrt{\textrm{Hz}}$ when excited with a continuous wave laser at 0.5~W.

These results pave the way for realizing a simple set-up magnetometer for precise single axis magnetic field measurements for example for accurate electric current measurements for stabilization purposes and magnetic communication applications.

\end{abstract}

\maketitle

\section{Introduction}

Nitrogen-vacancy centers in diamond have come to be recognized as versatile and powerful sensors for magnetic fields~\cite{Barry2020RMP, Zhang2022Frontiers}. Being part of a solid-state system, they are robust, easily scalable to the required sensitivity and adaptable to a number of applications using different diamond parameters and readout techniques.

Since a current flowing in a wire produces a magnetic field proportional to the strength of the current, NV sensors lend themselves as a non-invasive, contactless tool for measuring and monitoring currents. They are small and can be located close to the wire, and they can be read out in a way that is insensitive to temperature variations, unlike a shunt resistor. In this work we describe a device that was constructed with this purpose in mind. Going further, since the same device is capable of sensing magnetic fields that are varying in time, we describe a potential application for realizing a form of communication detecting magnetic signals at various distances.

The aim of this study was to construct a compact and portable tabletop experiment for high-sensitivity NV diamond based magnetic field measurements with an application for sensing magnetic fields caused by an electrical current.

NV centers in diamond are point-like defects in the carbon lattice~\cite{Doherty2013PhysRep,Jelezko2006PSSA}. For an NV center a vacancy (V) is surrounded by three carbon atoms and one carbon atom substituting nitrogen atom (N) that can be located in any of the four possible positions. The direction determined by the vacancy and the nitrogen atom is called the NV axis. There are four possible coexisting NV axis directions in any type of conventional diamond with NV centers in bulk.

NV centers in diamond can be used as a high sensitivity, contactless, non-invasive quantum sensor for: single axis magnetic field measurements~\cite{Fescenko2020} for example as a feedback loop for magnetic process stabilization purposes, 3D magnetic field measurements (50 pT/$\sqrt{\mathrm{Hz}}$~\cite{schloss_simultaneous_2018}) for GNSS denied magnetogeoreferencing, magnetic field imaging~\cite{Berzins2022} for creating magnetic field distribution images of soil samples, radio-frequency spectral analysis~\cite{Ludovic2018} for detecting radar signals, magnetic communication in regular RF damped situations (for example in caves, underground or underwater)~\cite{Morag2017optimization}.

\section{Experimental setup}
\begin{figure}
  \begin{center}
    \includegraphics[width=0.49\textwidth,trim={0pt 0pt 0pt 0pt},clip]{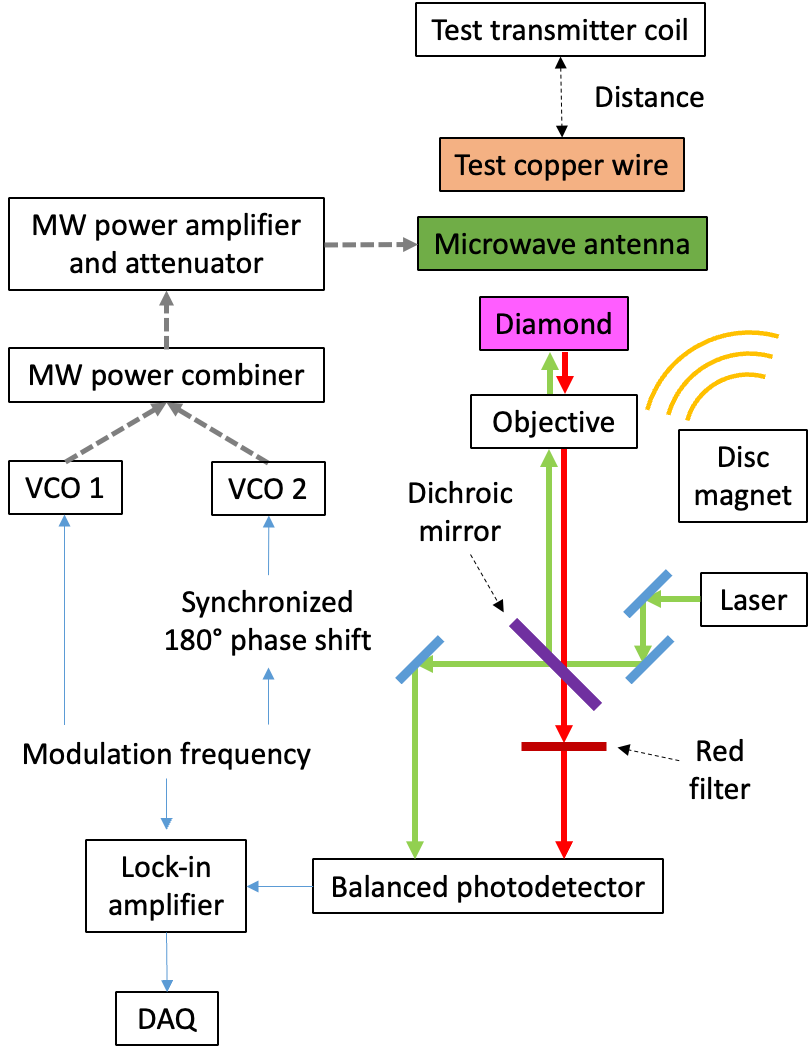}
  \end{center}
  \caption{Experimental setup. The microwave frequency was set by changing the voltage that was sent to the VCOs.}
  \label{experimental_setup}
\end{figure}

We constructed a tabletop ODMR set-up and applied a dual-resonance measurement approach~\cite{Fescenko2020} as shown in the block diagram in FIG.~\ref{experimental_setup} based on the experimental setup described in a previous research paper~\cite{Jani_2025}. 
We used a diamond (CVD) cut in the (110) plane. The NV$^-$ concentration
in the diamond sample was around 1~ppm and the hyperfine
structure line width was approximately 1~MHz~\cite{PhysRevB.111.064113}.

To excite the NV centers in the diamond, a 532~nm green laser (Cobolt Samba) operated at 500~mW output power was used. A 0.1 neutral density reflective filter was placed in front of the laser. The filter was placed at a slight angle to divert away any back reflections from the laser output aperture. Free-space mirrors were used to guide the light to the diamond. Mirrors were chosen instead of optical fibers, because they were found to be less prone to introducing fluctuations in the laser power delivered to the diamond. The green light was focused on the diamond adhered (using Electron Microscopy Sciences Crystalbond 509) to a printed circuit board, which implemented a microwave (PCB-MW) antenna~\cite{Sasaki2016}, through an objective (Thorlabs RMS40X, NA 0.65, WD 0.6~mm). The emitted red fluorescence from the diamond was collected through the same objective. A dichroic mirror (Thorlabs DMLP550) and a long pass filter (Thorlabs FELH0550) was used to separate the emitted red fluorescence from the exciting green light and was guided to the first channel of a balanced photodetector (Thorlabs PDB210A/M). The second channel of the balanced photodetector was used to monitor the laser power. In this way, any noise caused by the laser power fluctuations was mitigated (the laser power monitoring channel was balanced using a continuously variable ND filter Thorlabs NDL-25C-2). The balanced output signal from the photodetector was used as an input to a lock-in amplifier (Anfatec 250~kHz 2-phase USB Lock-In Amplifier).

A reference frequency for the lock-in amplifier of 4~kHz was supplied by a RedPitaya STEMlab 125-14 board that also supplied two 180$^{\circ}$ phase shifted sine signals ($\pm 1$~V) that were used to drive two voltage controlled oscillators (VCOs, Crystek CVCO55BE-2560-3200) with the use of operational amplifiers (TL072CP) to reach the necessary voltages (generate the necessary microwave frequencies) to perform the dual resonance approach in ODMR measurements~\cite{Fescenko2020}. In this approach, both of the NV center's ground state transitions $\vert m_S = 0 \longrightarrow m_S = -1\rangle$ and $\vert m_S = 0 \longrightarrow m_S = +1\rangle$ are excited at the same time leading to a higher degree of population transfer between the energy levels, which provides a higher lock-in amplifier output amplitude compared to a single resonance approach. This technique is not sensitive to temperature variations of the diamond, at least to first order.

Both transitions experience identical shifts of the zero-field splitting parameter $D$, the dual-resonance scheme suppresses common-mode temperature-induced frequency drifts while retaining sensitivity to the Zeeman splitting~\cite{Fescenko2020}. The lock-in amplifier output in this experimental configuration is directly proportional to the difference (due to the 180$^{\circ}$ phase difference) between the two MW transition energies $f_{-1}$ and $f_{+1}$ in the frequency domain:
\begin{equation}
\begin{split}
f_{-1} &= D(\Delta T) - 1\cdot\gamma B ,\\
f_{+1} &= D(\Delta T) + 1\cdot\gamma B ,\\
f_{+1}-f_{-1} &= 2\gamma B ,\\
B &= \frac{f_{+1}-f_{-1}}{2\gamma} ,
\end{split}
\label{delta_f}
\end{equation}
where $\gamma = 2.803$~MHz/G is the electron gyromagnetic ratio and $B$ is the external axial magnetic field that the NV centers are sensing. Thus, the lock-in amplifier extracts only the magnetic field amplitude, and eliminates the shifts caused due to temperature changes $\Delta T$ associated with the zero field parameter D=2870~MHz equal to 75~kHz/K~\cite{acosta_temperature_2010}.

A Minicircuits power splitter/combiner (ZX10-2-183-S+) was used to combine the outputs from both of the VCOs, along with, a Minicircuits MW amplifier (ZHL-1W-63-S+ and a 10 dB attenuator) to amplify and deliver the microwaves to the PCB-MW antenna \cite{Sasaki2016}. A MCC USB-1608GX-2AO-OEM DAQ running at a sampling rate of 100~kS/s was used to record the data from the lock-in amplifier output that was then interpreted as the measured magnetic field.

A permanent 6~cm diameter disc magnet was used to generate a bias magnetic field of $\approx$~0.83~mT at the location of the excitation spot on the diamond. The disc magnet was positioned so that one of the NV axis in the diamond is parallel to the magnetic field generated by the magnet (by aligning the ODMR peaks as shown in FIG.~\ref{fig:0A_ODMR}).

Two types of test measurements were performed:
\begin{itemize}
    \item monitoring how the magnetic field generated by an electrical current flowing through a test wire (the diameter of the wire was 5.5~mm and the distance from the diamond sample to the surface of the wire was 3~mm) changes over time, a high-termal-capacity wire wound power resistor of 0.1~$\Omega$ (HSC300R10J) was used as a load resistor in series with the test wire;
    \item detecting low frequency magnetic field signals at larger distances using a test coil as a magnetic field "transmitter" (the coil parameters are described in the later sections).
\end{itemize}

\section{Sensing electrical current changes via short distance contactless magnetic field sensing}

The first stage of testing the performance of the experimental device was to measure magnetic fields that are generated by an electrical current of various strengths flowing through a test wire close to the diamond.

The experimental setup allows for monitoring of electrical current deviations around any chosen operating point, provided that the ODMR spectrum is acquired under approximately the same magnetic field conditions.

\begin{figure}
  \begin{center}
    \includegraphics[width=0.49\textwidth,clip]{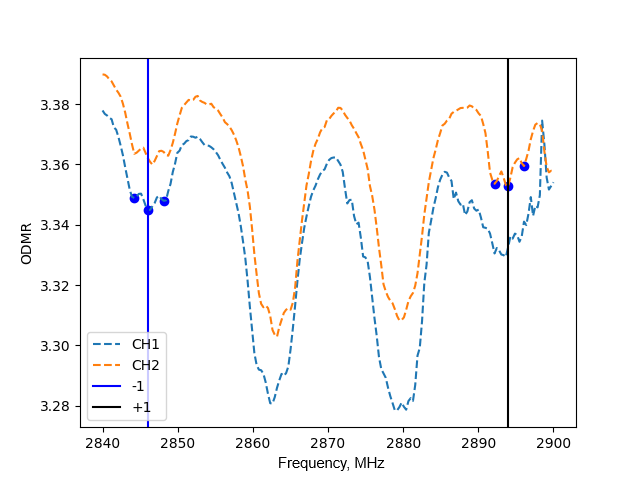}
  \end{center}
  \caption{ODMR spectrum with no electrical current in the test wire. The blue curve shows some noise in the $\vert m_S = 0 \longrightarrow m_S = +1\rangle$ transition frequency range, this is attributed to the quality and performance of the VCOs and the electrical boards that they were mounted on that were used in the experiment.}
  \label{fig:0A_ODMR}
\end{figure}

We performed measurements at low electrical currents (0 - 0.4~A) and for a higher current of 50~A demonstrating the versatility of such a device. A typical ODMR spectra for when there is no additional magnetic field other than the ambient bias field that the diamond is exposed to is shown in FIG.~\ref{fig:0A_ODMR}.

\begin{figure}
  \begin{center}
    \includegraphics[width=0.49\textwidth,clip]{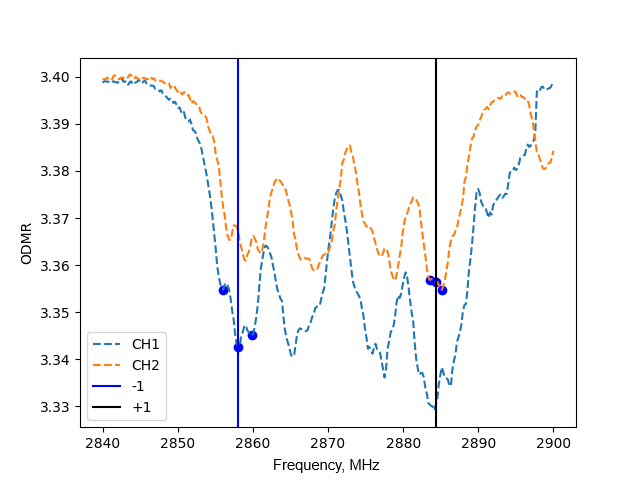}
  \end{center}
  \caption{ODMR spectrum with an electrical current of 50~A in the test wire. Distortions in the ODMR spectra compared to the 0~A situation can be attributed to such effects as thermal changes due to rapid heating of the test wire from the electric current and as the bias field is partially formed by the current, it is also dependent on the current stability (increased magnetic-field inhomogeneity across the illuminated NV ensemble). The limits of the VCO operating range and efficiency also impact this.}
  \label{fig:50A_ODMR}
\end{figure}

The ODMR spectrum with a 50~A current applied to the test wire is shown in FIG.~\ref{fig:50A_ODMR}. It can be seen how the higher electrical current that is changing in time affects the measured ODMR signal. Nevertheless, the ODMR signal was sufficient enough for performing magnetic field measurements using the dual-resonance method.

The precise frequencies for both of the NV center's ground state transitions $\vert m_S = 0 \longrightarrow m_S = -1\rangle$ and $\vert m_S = 0 \longrightarrow m_S = +1\rangle$ were extracted from the measured ODMR spectra and used for the dual-resonance measurements of the magnetic field, selecting the central hyperfine component for each transition.

\begin{figure}
  \begin{center}
    \includegraphics[width=0.49\textwidth,clip]{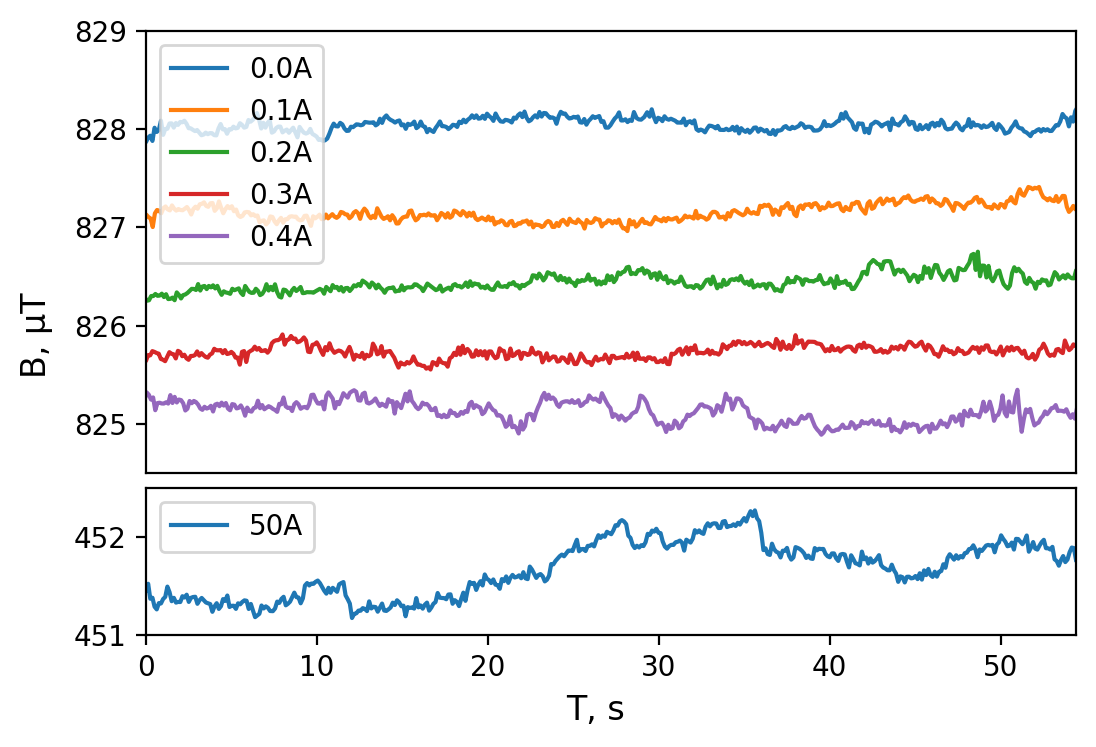}
  \end{center}
  \caption{Magnetic field over time (in seconds) for multiple electrical current values.}
  \label{fig:current}
\end{figure}

Magnetic field fluctuations were monitored for low electrical current and for the higher 50~A current as shown in FIG.~\ref{fig:current}. Each data point was obtained every 125~ms.

The standard deviation of the measured magnetic field was found to be 281.3~nT at 50~A, compared to 85.2~nT at zero current, see TABLE~\ref{tab:B_statistics}. A possible reason for the larger changes in the detected magnetic field for the higher electrical current can be the fact that the large current heats up the test wire and that causes additional changes in the detected magnetic field, the changes of temperature can also cause small mechanical movements of the test wire with respect to the diamond.
\begin{table}[t]
\centering
\caption{Mean magnetic field and standard deviation measured over time for different applied currents.}
\label{tab:B_statistics}
\begin{tabular}{| c | c | c |}
\hline
Current $I$ (A) & $\langle B \rangle$ ($\mu$T) & $\sigma_B$ (nT) \\
\hline
0.0  & 828.06 & 85.2 \\
0.1  & 827.17 & 92.6 \\
0.2  & 826.47 & 114.7 \\
0.3  & 825.75 & 78.4 \\
0.4  & 825.14 & 109.6 \\
50   & 451.67 & 281.3 \\
\hline
\end{tabular}
\end{table}

From these results we can calculate electrical current to magnetic field conversion factors. If we use the low current range then the factor is $\alpha = (\langle B \rangle _{I_0} - \langle B \rangle _{I_1})/(I_1-I_0)=(828.06-825.14)/(0.4-0.0)=7.3 \mu$~T/A while using the 50~A data point we obtain $7.5 \mu$~T/A.
If we use this value with the average value of the standard deviations for the low electrical currents $\sigma_B=96$~nT, we can deduce that the smallest change in electrical current that we would be able to objectively detect is $\sigma_B/\alpha=13.1$~mA.

\section{Sensing low frequency magnetic field remote signals}

Based on the results in the previous section, we moved on to the next stage of testing the performance of the experimental device by detecting low frequency magnetic field signals at larger distances, up to 10~m in this case.

The experimental setup (the NV center based magnetic field sensor) was used as a proof-of-concept receiver for low-frequency magnetic field communication. Unlike conventional radio-frequency communication (utilizing the electric field), this approach relies on time varying magnetic fields at frequencies well below the RF regime, making it suitable for environments where electric field propagation is strongly attenuated~\cite{Morag2017optimization}.

\begin{figure}
  \begin{center}
    \includegraphics[width=0.49\textwidth,trim={0pt 0pt 0pt 20pt},clip]{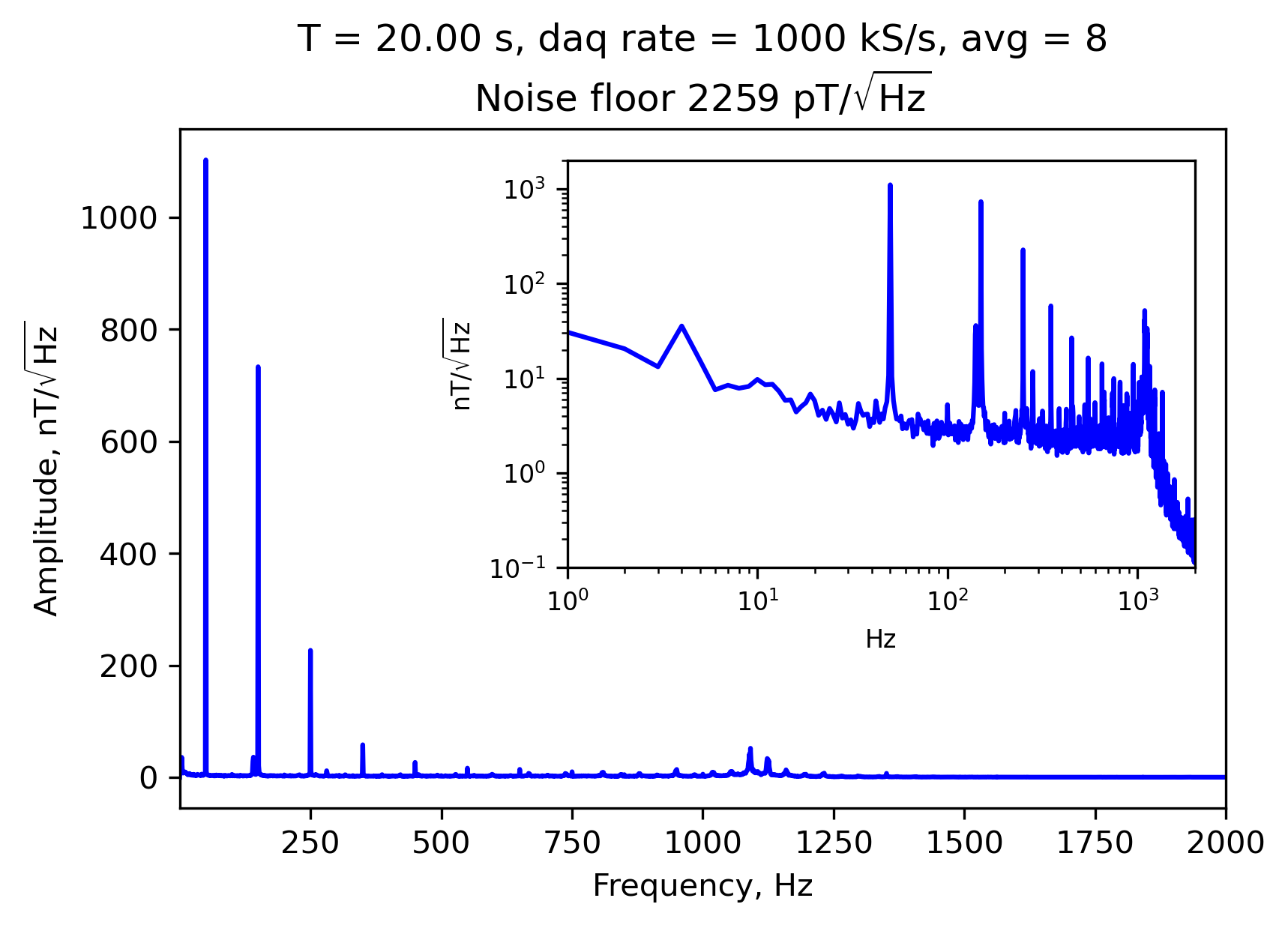}
  \end{center}
  \caption{Typical noise-floor (2.26~nT/$\sqrt{\textrm{Hz}}$) spectra with no additional time varying magnetic field showing signals at 50~Hz (from the surrounding electronics) and its harmonics. The time constant (1~ms) in combination with the roll-off settings (24~dB/oct) for the lock-in amplifier limits the actual sensing range to slightly below 1 kHz as higher frequency signals are increasing being filtered out from the measured magnetic field (see inset with log-log axis).}
  \label{fig:noisefloor}
\end{figure}

In order to do so we used the power spectral density (PSD) analysis on the measured magnetic field over time~\cite{Jani_2025} to determine the noise-floor of our magnetic field measurements, see FIG.~\ref{fig:noisefloor}. The noise-floor spectra were used to detect and monitor magnetic fields varying at specific frequencies.

With this in mind we made a portable test coil (see details in TABLE~\ref{coil}) system and performed measurements at various distances while generating a time varying magnetic field with a frequency of 120~Hz (chosen not to coincide with 100~Hz or 150~Hz signals coming from the electronics).

\begin{figure}
  \begin{center}
    \includegraphics[width=0.49\textwidth,trim={0pt 0pt 0pt 0pt},clip]{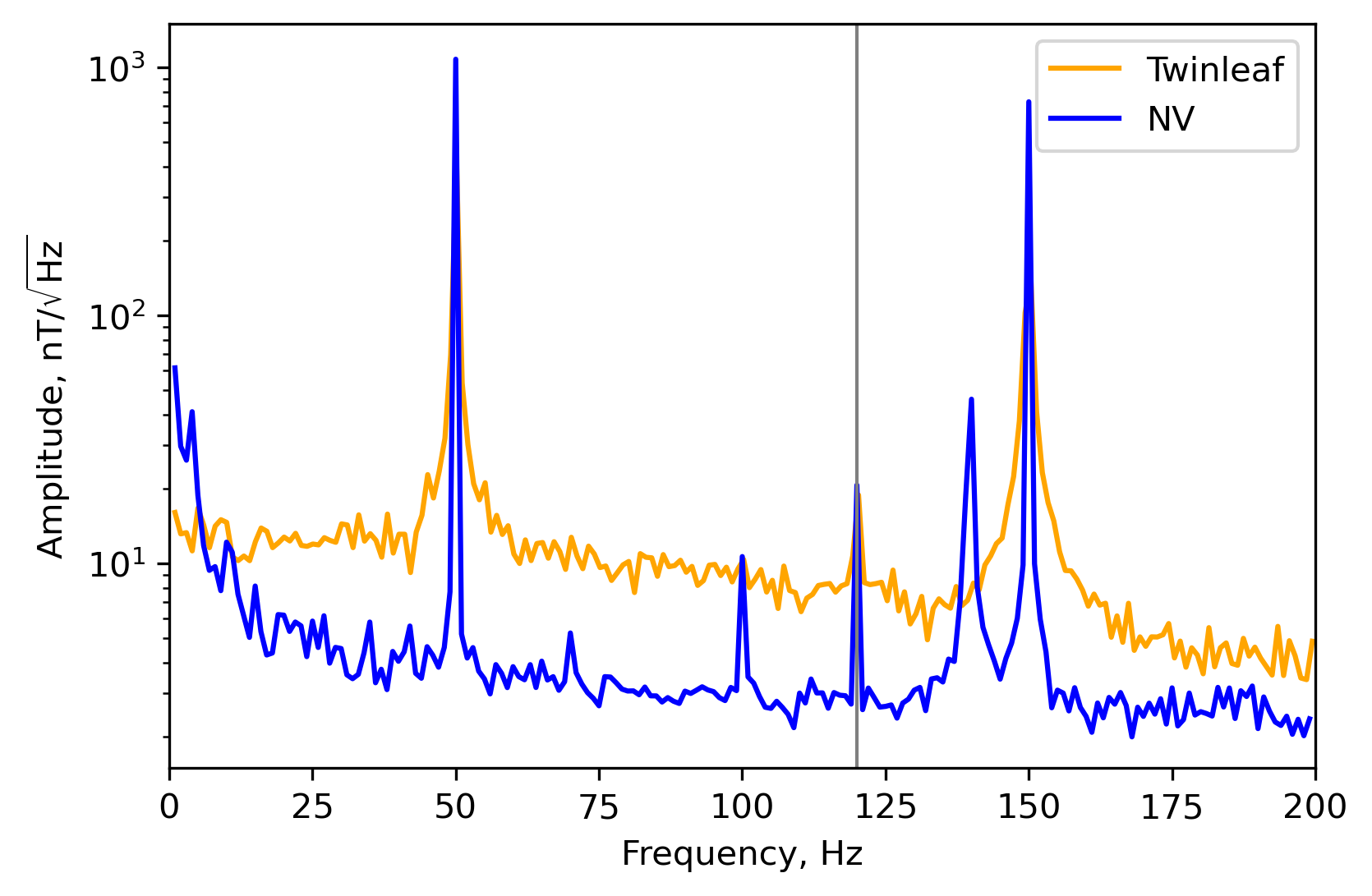}
  \end{center}
  \caption{Noise-floors for the NV based sensor and the Twinleaf sensor showing a detected magnetic field signal at 120~Hz (indicated by the grey vertical line) at a distance of 8~m. It can be seen that in the frequency range from 0 to 200~Hz (limited by the measurement rate of the Twinleaf sensor) the noise floor of the NV sensor is lower.}
  \label{fig:noisefloor_both}
\end{figure}

\begin{figure}
  \begin{center}
    \includegraphics[width=0.49\textwidth,clip]{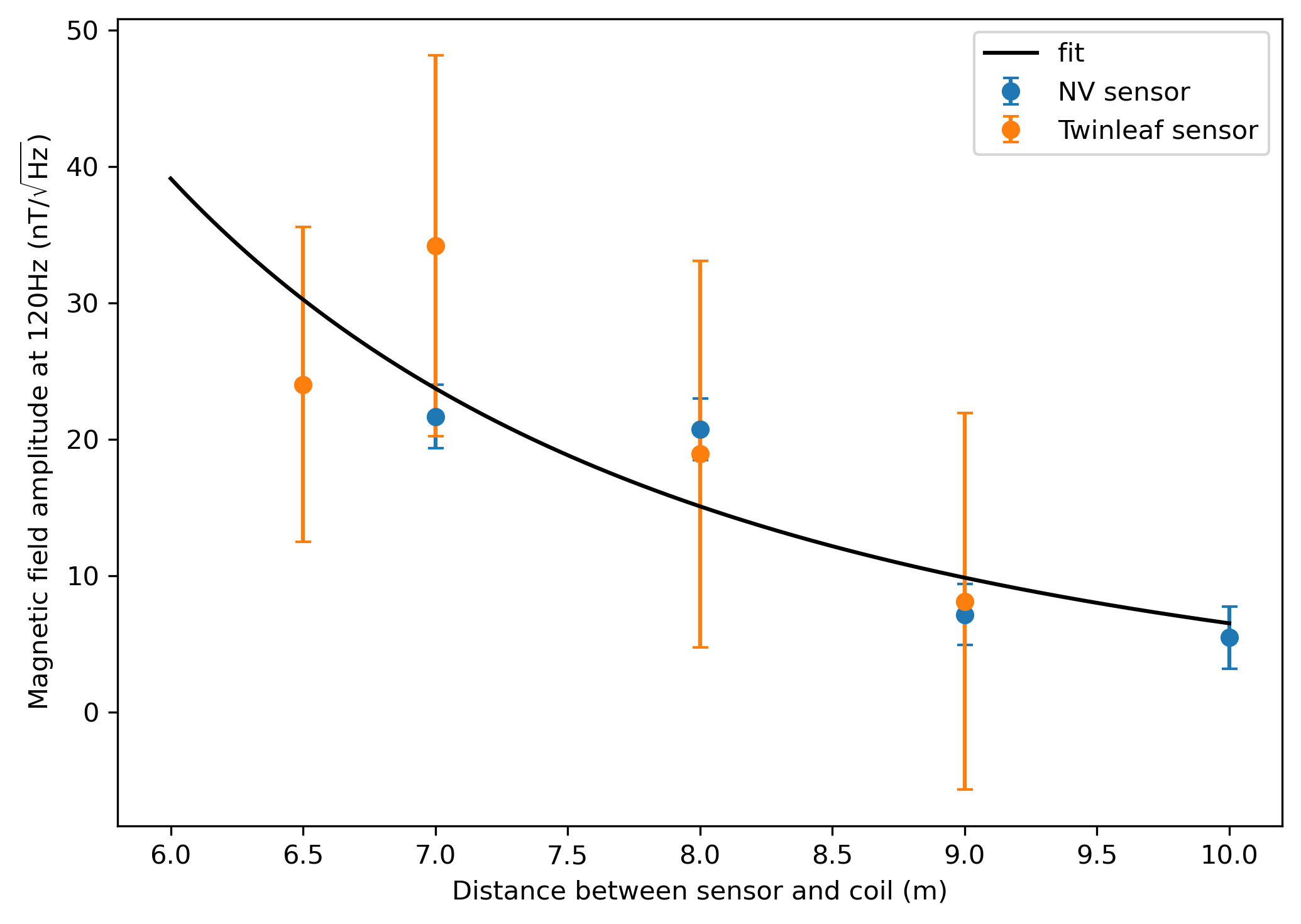}
  \end{center}
      \caption{Measured magnetic field intensity of the 120~Hz signal over multiple distances. Due to the noise-floor of the Twinleaf sensor measurements beyond 9 meters did not detect the magnetic field signal at 120~Hz. The error bars for the NV sensor are approximately 5.8 times smaller than those for the Twinleaf sensor.}
  \label{fig:distances}
\end{figure}

In order to validate and benchmark our NV-based magnetic field sensor we performed magnetic field measurements using a magnetoresistive commercially available sensor from Twinleaf (VMR Magnetoresistive Vector Magnetometer) as well. The dependence of the received magnetic field strength at 120~Hz (see FIG.~\ref{fig:noisefloor_both}) on the distance for both magnetic field sensors is shown in FIG.~\ref{fig:distances}.

The uncertainties in FIG.~\ref{fig:distances} were estimated by combining the measured noise-floor values with an additional contribution arising from the distance uncertainty, assumed to be 5~cm. All data points were included in a weighted least-squares fit using the following model:
\begin{equation}
    B(d)=\frac{C}{(d^2+R^2)^{3/2}}+y_0 ,
\end{equation}
where $d$ is the distance between the test coil and the magnetic field sensor, $R$ is the radius of the test coil while $C$ and $y_0$ are the fit parameters. This is basically the analytical formula for calculating the coil-induced magnetic field \ref{eq:B_coil}.

\begin{figure}
  \begin{center}
    \includegraphics[width=0.49\textwidth,clip]{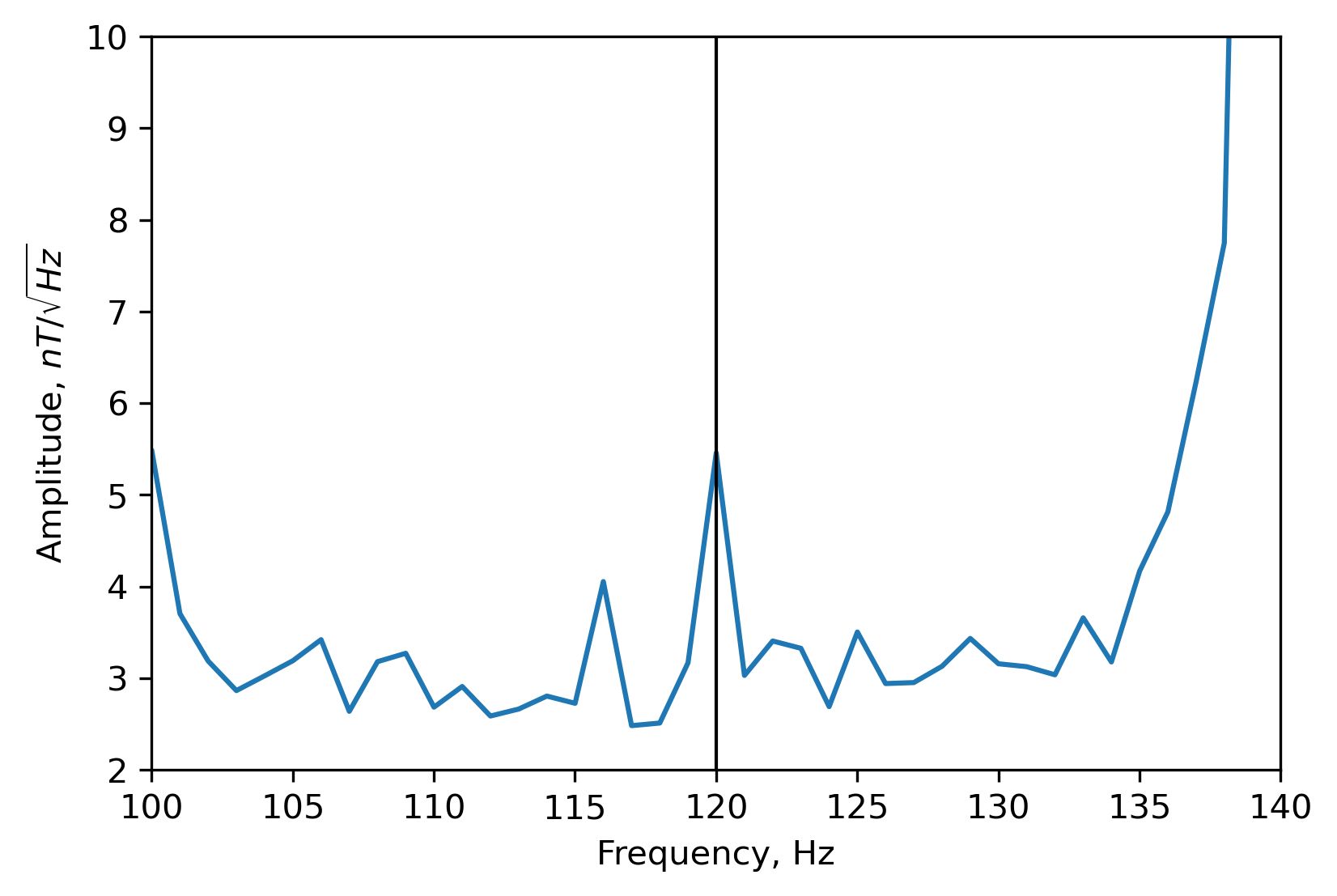}
  \end{center}
      \caption{Detection of 120~Hz signal over a distance of 10~m.}
  \label{fig:10m}
\end{figure}

Figure \ref{fig:10m} shows the successful reception of a 120~Hz magnetic field signal over a distance of 10~m. This demonstrates that low-frequency magnetic communication using NV-based magnetometers is feasible over meter-scale distances, even at frequencies where conventional antennas would be impractically large~\cite{Chen2020ultra, FANG2025100900}.

To test for benefits in using magnetic field for communication purposes instead of electric field we performed measurements using a 50~L volume of water between the test coil and the magnetic field sensors, close to the magnetic field sensors. The idea being that the water should not affect the magnetic field measurements much and that the electric field should be attenuated in the water.

\begin{figure}
  \begin{center}
    \includegraphics[width=0.49\textwidth,clip]{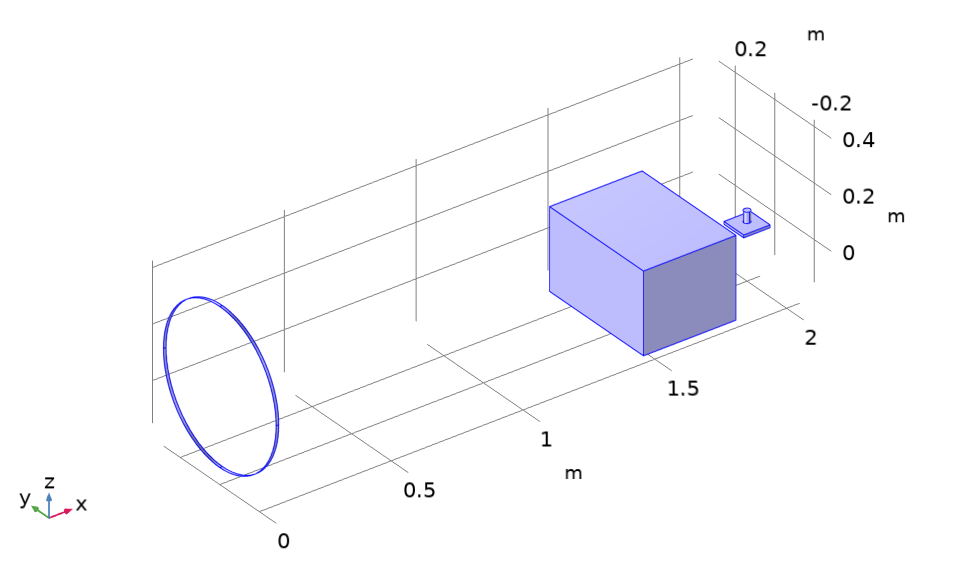}
  \end{center}
      \caption{Numerical model setup containing coil, water container and NV center sensor table.}
  \label{fig:model_setup}
\end{figure}

To verify the experimental data we performed numerical calculations. Modeling was done using the COMSOL Multiphysics finite element solver. The model geometry comprises a coil, water container and NV center sensor, all enclosed in an air domain, see FIG.~\ref{fig:model_setup}. A total of 300k mesh elements were used. The coil input parameters are shown in TABLE~\ref{coil}. The coil-induced magnetic field decays to zero at infinity thanks to the infinite element domain boundary condition applied to the outer layer of the spherical air domain. A container filled with tap water ($\sigma = 0.0137$ S/m, $\varepsilon_r = 80$) is placed in front of the magnetic field sensor to measure any signal attenuation. A time-harmonic solution is obtained using the Geometric multigrid solver.

\begin{table}[h!]
\caption{Coil parameters.\label{coil}}
\centering
 \begin{tabular}{|c | c|}
 \hline
 Parameter, symbol(unit) & Value \\ [0.5ex] 
 \hline
 Voltage, $V$ (V) & 10 \\
 Impedance, $Z$ ($\Omega$) & 50 \\
 Current, $I$ (A) & 0.2 \\
 Frequency, $f$ (Hz) & 120 \\
 Inductance, $L$ (H) & 0.06 \\
 Diameter, $D_{coil}$ (m) & 0.57 \\
 Thickness, $h$ (m) & 0.01 \\
 Wire diameter, $d_w$ (m) & 0.00079 \\
 Turns, $N$ (-) & 191 \\ [0.1ex]
 \hline
 \end{tabular}
\end{table}

\begin{figure}
  \begin{center}
    \includegraphics[width=0.49\textwidth,clip]{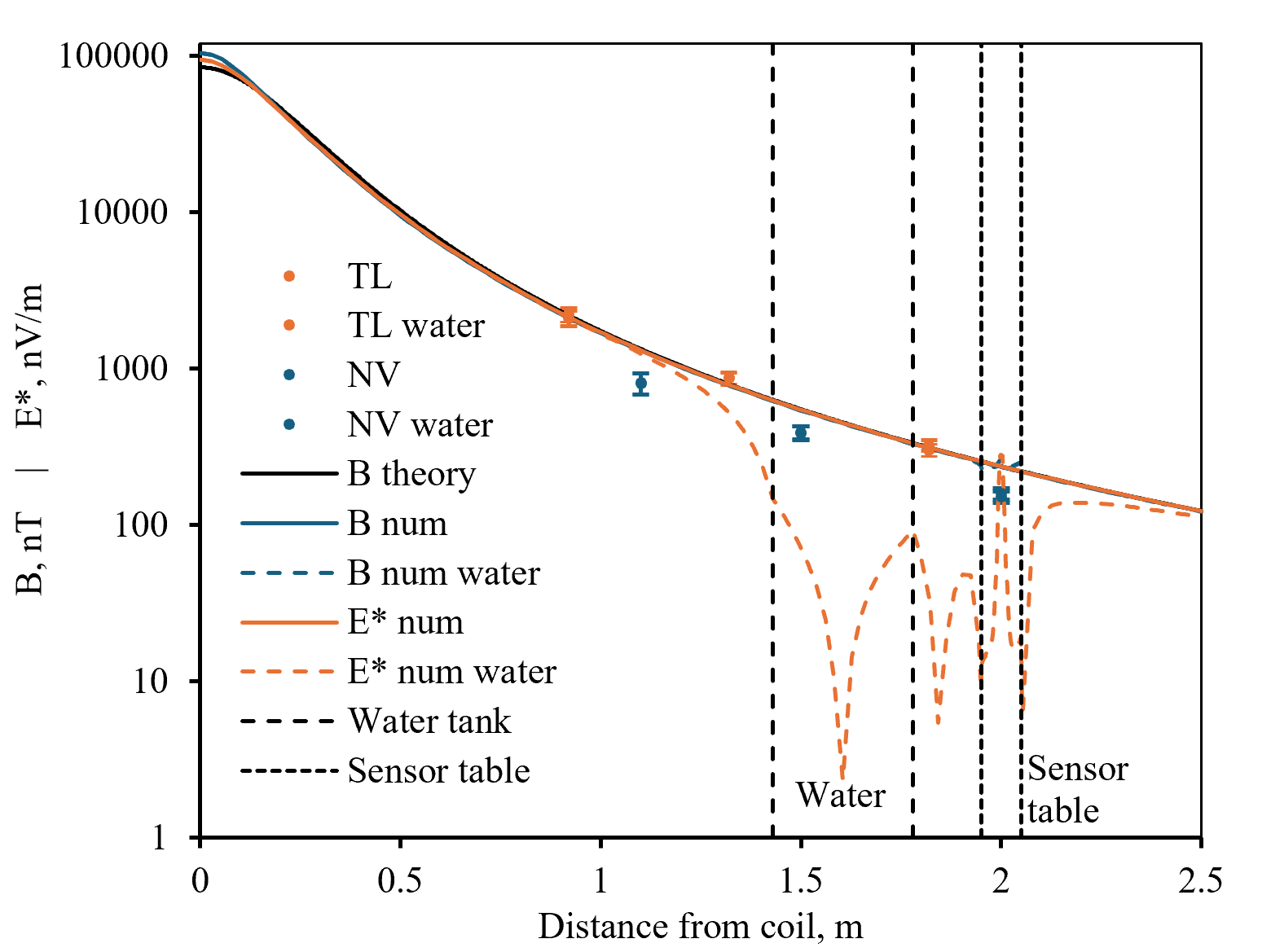}
  \end{center}
      \caption{Simulation comparison with theory and experiment. The vertical lines show the position of the water container and sensor table.}
  \label{fig:model_results}
\end{figure}

The numerical results are shown in FIG.~\ref{fig:model_results}. The analytical solution for the coil-induced magnetic field is included for validation.
\begin{equation}
    B(x)=\frac{\mu_0}{4\pi}\frac{N I R\cdot2\pi R}{(x^2+R^2)^{3/2}} ,
\label{eq:B_coil}
\end{equation}
where $\mu_0,~R,~x$ are vacuum magnetic permeability, coil radius, and distance from coil, respectively. The slight mismatch between the analytical and numerical results near the coil is caused by the axis offset of 0.14~m to replicate the experimental setup. Despite this, there is excellent agreement between the analytical, numerical and experimental results. The difference between the Twinleaf and the NV center sensor measurements is beyond the error bars, which could potentially be explained by an offset optimal reading angle of the NV center sensor (the NV axis direction in the diamond that is used to sense the magnetic field with respect to the applied test magnetic field). In the experiment, an optimal sensor alignment was found after scanning angles in the horizontal plane. Vertical alignment was not performed. We hypothesize that there could even be a maximum of 50-degree misalignment in the vertical plane.

From FIG.~\ref{fig:model_results} we observe that the magnetic field is not impacted by the water container and only slightly affected by the sensor's aluminium table platform. However, the electrical field is affected by the presence of water as indicated by the graph line oscillations and deviation from the void case. Note that the electrical field has been scaled down, $E^*=E/52.5$, to match the vertical axis values of the magnetic field. Absolute values for the electric field here do not matter as we are capturing trends and relative changes. Here, the electrical field decay follows the trend of $1/Distance^3$. The same tendency is displayed by the coil-induced low frequency magnetic field decay.

\section{Sensing information bits encoded in low frequency magnetic field signals}

To test potential practical communication possibilities information was encoded using discrete frequency modulation of the magnetic field generated by the transmitting coil. Each distinct frequency in the range 100–150~Hz was attributed as an information bit, effectively implementing a low-frequency frequency-shift keying (FSK) scheme~\cite{Hott2021MRFSK}. To improve robustness against noise, the transmitted data was encoded using a Hamming(7,4) error-correcting code~\cite{Hamming1950}, with encoded symbols mapped to eleven distinct frequencies representing an ASCII character.

\begin{figure}
  \begin{center}
    \includegraphics[width=0.49\textwidth,clip]{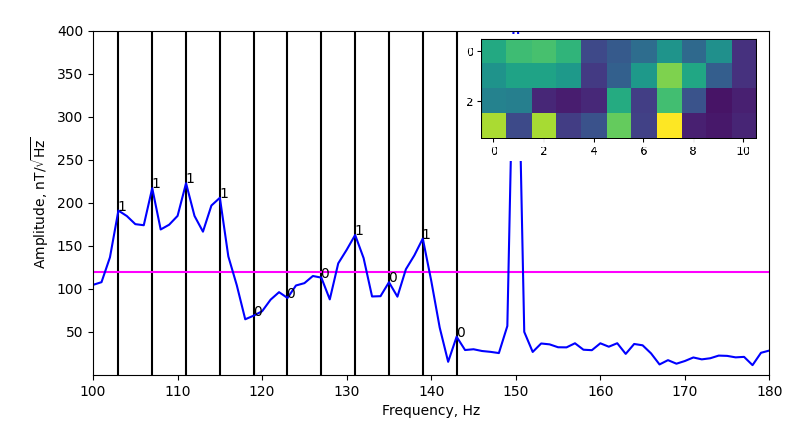}
  \end{center}
  \caption{Example for the transmission and reception using modulated magnetic field. The graph shows the last received symbol "t" as bits encoded in various magnetic field frequencies that reach above a certain threshold. The inset on the right side shows the received bits as a "waterfall" type diagram, the top row showing the latest received symbol and from bottom to top can be decoded as the transmitted message "Test". The large peak at 150~Hz is attributed to harmonics of 50~Hz caused by electronics surrounding the diamond. The detection of these signals was done at a distance of 2~m.}
  \label{fig:transmission}
\end{figure}

An example transmission is shown in FIG.~\ref{fig:transmission}. The receiver (the NV based magnetic field sensor in this case) reconstructs the transmitted sequence by identifying spectral peaks in successive averaging windows. The image shows the frequency-domain representation of the received signal, with vertical markers indicating the predefined transmission frequencies; the presence or absence of a spectral peak at each frequency corresponds to logical "1" or "0", respectively. The inset shows the temporal history of the detected symbols by bits, where the top row corresponds to the most recent reception window. The correspondence between transmitted and received frequencies demonstrates reliable symbol detection under the applied experimental conditions.

Synchronization between the transmitter and receiver was achieved without any shared clock or external triggering. A single initialization sweep through all available transmission frequencies was used as a preamble, during which the receiver determined the temporal overlap between its averaging window and the transmitted signal sequence.

\section{Results and conclusions}

\begin{figure}[h]
  \begin{center}
    \includegraphics[width=0.49\textwidth,clip]{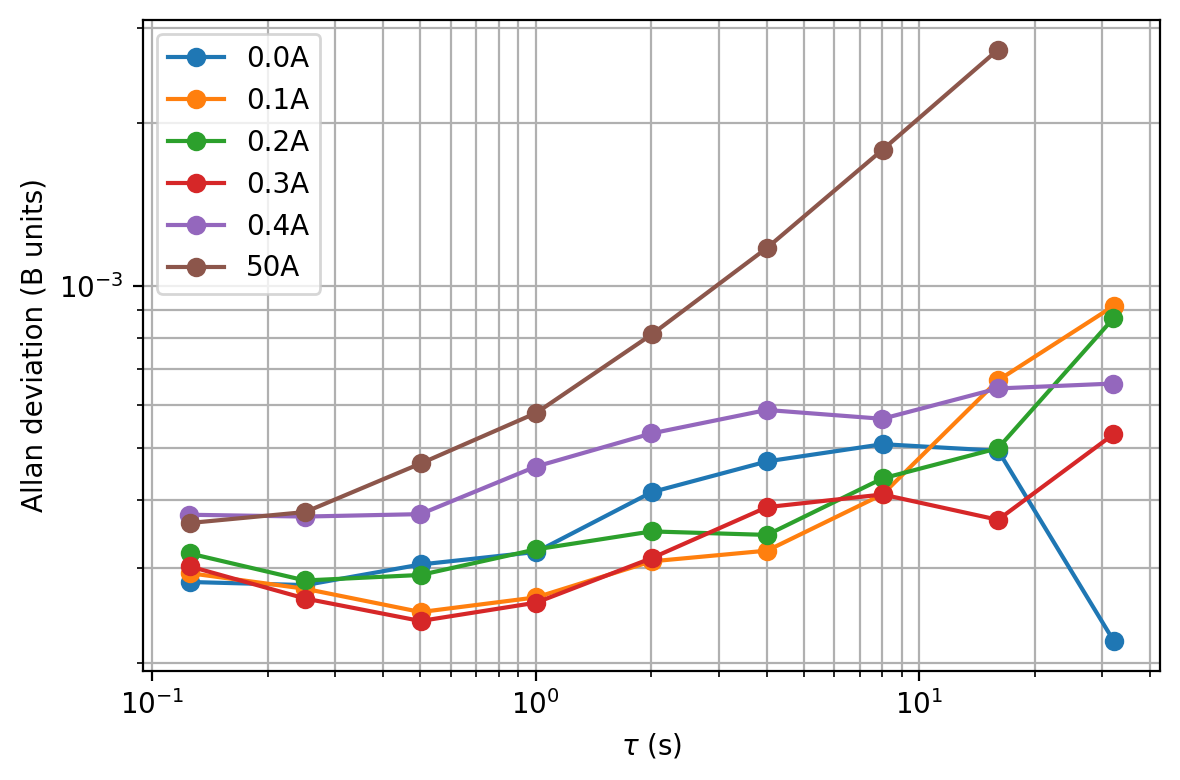}
  \end{center}
  \caption{Allan deviation for the magnetic field next to the electrical current test wire.}
  \label{fig:allan}
\end{figure}

To characterize the long-term stability of the magnetic field measurements, an Allan deviation analysis was performed for the various electrical current measurements; the resulting log-log plots are presented in FIG.~\ref{fig:allan}. For all low-current settings, the Allan deviation exhibits a near-zero slope at short averaging times, indicating flicker-dominated magnetic field noise~\cite{Allan1987}. For higher electrical current, the Allan deviation transitions to a positive slope approaching +0.5 in the log-log plot, consistent with random-walk behavior likely caused by the specific current-source drift and thermal fluctuations in the test wire.

To estimate what the theoretically best magnetic field sensitivity could be achievable in an ideal case using the ODMR method the shot-noise-limited magnetic field sensitivity was calculated to be $585~\mathrm{pT}/\sqrt{\mathrm{Hz}}$. The calculation was done using the following relation:
\begin{equation}
\eta = \frac{4}{3\sqrt{3}}\frac{h}{g_e \mu_B}\frac{\gamma}{C}\sqrt{\frac{G \, h \, c}{V \, \lambda}} ,
\label{shot-noise}
\end{equation}
where $\eta$ denotes the shot-noise-limited magnetic field sensitivity, $h$ is Planck’s constant, $g_e$ - the electron $g$-factor, $\mu_B$ - the Bohr magneton, $\gamma$ - the ODMR linewidth, $C$ - the ODMR contrast, $G$ - the photodetector gain (V/W), $V$ - the detected photovoltage, $\lambda$ - the optical wavelength, and $c$ - the speed of light in vacuum~\cite{Barry2020RMP}. In this particular case we used the central hyperfine component of the $\vert m_S = 0 \longrightarrow m_S = -1\rangle$ transition in the ODMR spectra shown in FIG.~\ref{fig:0A_ODMR} to get the linewidth, contrast and the photovoltage for the calculation (the parameters were 1.1~MHz, 0.94~\% and 3.3~V respectively).

Based on the results in TABLE~\ref{tab:B_statistics}, the calculated electrical current resolution and the estimated shot-noise-limited magnetic field sensitivity~\eqref{shot-noise} it can be seen that the resolution for detecting changes in the electrical current for the close distance measurements could, in principle, be improved approximately 164 times (96~nT~/~0.585~nT) to 13.1~mA~/~164~$\approx$~80~$\mu$A (using the shot-noise-limited value as the magnetic field measurement accuracy -- shot-noise-limited magnetic field detection). Alignment of the diamond with respect to the test wire so that the magnetic field sensing NV axis is perpendicular to the test wire longitudinal direction would further improve the sensitivity to changes in the magnetic field form the wire.

The noise-floor values extracted from the PSD spectra are significantly lower than the standard deviation values obtained from the time-domain magnetic field measurements. This difference arises because the PSD characterizes noise within a certain, limited frequency band, while the time-domain standard deviation integrates noise contributions over the entire frequency spectrum, including low-frequency drift and flicker noise. As confirmed by the Allan deviation analysis, the present system is dominated by low-frequency noise processes, which have a negligible contribution to the PSD noise-floor but strongly affect long-term stability.

Looking at FIG.~\ref{fig:distances} and FIG.~\ref{fig:model_results} it can be seen that for the larger distances the trend for the NV measurements to be smaller in value compared to the Twinleaf sensor (explained by an angle difference between the detected magnetic field and the NV axis in the diamond) is not as prominent. One explanation for this can be the fact that the transmitted magnetic signal at the larger distances is closer to the noise-floor value and thus more prone to be lost in the noise-floor.

Compared to previous results in terms of magnetic field sensitivity and noise-floor~\cite{Jani_2025} of a similar, but much more larger, power consuming and expensive, experimental setup it can be seen that the quality of microwaves~\cite{PhysRevResearch.6.043148} is a significant factor (laboratory grade MW generators vs compact VCOs).

The present implementation, although compact, as was the goal of this study, is limited by bias magnetic field inhomogeneity, laser power noise, and microwave noise, which collectively prevent reaching the shot-noise-limited sensitivity. Improvements are expected from optimized microwave delivery, potentially using a phase-locked loop (PLL) solution in combination with a VCO (yielding a more stable MW output) and active temperature stabilization.

These results support the feasibility of low-frequency magnetic field based communication using magnetometers (NV-center based and others), complementing earlier work on magnetic induction communication in RF-attenuating environments~\cite{Morag2017optimization} - under rocks, in caves, underwater, in specific urban areas. Higher distances for detecting time varying magnetic field signals could be achieved using a more powerful transmitter coil and a lower noise-floor magnetic field sensor.

\section*{Acknowledgments}
We acknowledge the support from the Latvian Council of Science, project No. lzp-2021/1-0379: “A novel solution for high magnetic field and high electric current stabilization using color centers in diamond”.

The used Cobolt laser was purchased by the funds provided by LLC “MikroTik” and the Foundation of the University of Latvia, project No. 2320: "A system for precise detection of double quantum magnetic resonance".

\newpage
\bibliography{main}
\bibliographystyle{apsrev}

\end{document}